# Efficient second-harmonic emission via strong modal overlap in single-resonant lithium niobate nanocavity

Zhi Jiang,[1] Danyang Yao,[1*] Yu Gao,[1] Xu Ran,[1] Duomao Li,[1] Erqi Zhang,[1] Jianguo Wang,[2] Xuetao Gan,[2*] Jinchuan Zhang,[3*] Fengqi Liu,[3] and Yue Hao[1]

[1]State Key Laboratory of Wide-Bandgap Semiconductor Devices and Integrated Technology, School of Microelectronics, Faculty of Integrated Circuits, Xidian University, Xi'an, 710071, China;

[2]Key Laboratory of Light Field Manipulation and Information Acquisition, Ministry of Industry and Information Technology, and Shaanxi Key Laboratory of Optical Information Technology, School of Physical Science and Technology, Northwestern Polytechnical University, Xi'an, 710129, China;

[3]Laboratory of Solid-State Optoelectronics Information Technology, Institute of Semiconductors, Chinese Academy of Sciences, Beijing, 100083, China;

* dyyao@xidian.edu.cn; xuetaogan@nwpu.edu.cn; zhangjinchuan@semi.ac.cn;

**ABSTRACT:**

High-efficiency second-harmonic generation (SHG) in compact integrated photonic systems is crucial for advancing nonlinear optical technologies. However, achieving exceptional conversion efficiencies while maintaining stable performance remains a significant challenge. Here, we report a high-Q single-resonant photonic crystal nanobeam cavity (PCNBC) on a polymer-loaded lithium niobate on insulator (LNOI) platform, which enables bright second-harmonic (SH) emission. Through synergistic optimization of modal confinement and spatial overlap in a y-cut LN architecture, our device achieves a normalized SHG conversion efficiency of 163%/W, outperforming previous LN-based photonic crystal cavities LN-based photonic crystal cavities by over three orders of magnitude. The visible SH emission at 768.77 nm exhibits a single-lobe radiation pattern with precise spectral alignment between fundamental (FH) and second-harmonic (SH) modes, a critical feature for integrated photonic circuits. Remarkably, the conversion efficiency remains stable under thermal variations up to 20°C, addressing a key limitation of multi-resonant systems. High-order cavity modes are directly visualized via CCD imaging, confirming strong spatial overlap. This work establishes a record SHG conversion efficiency for LN microcavities and provides a scalable, temperature-insensitive architecture for nonlinear light sources, with immediate applications in quantum optics and chip-scale interconnects.

## ■ INTRODUCTION

Integrated coherent light sources remain indispensable yet challenging components for modern photonics[1]. Nonlinear optical processes such as second-harmonic generation (SHG) offer viable pathways toward chip-scale sources, with applications spanning sensing,[3] frequency combs,[4,5] nonlinear electro-optic modulators,[6,7] all-optical communications,[8,9] and quantum optics.[10,11]. Enhancing these processes necessitates maximizing both energy density[12] and spatial mode overlap[13], thereby driving innovations in nanophotonic cavity design.

In high-Q resonant cavities, SHG conversion efficiency scales with $Q^2/V$, where Q is the cavity quality factor and V is the modal volume.[14,15] Photonic crystal (PhC) nanocavities are particularly effective at confining light in small mode volumes while maintaining high Q factors,[15] making them ideal for amplifying nonlinear effects and enabling integration with resonators and waveguides.[16-18] Theoretical studies predict substantial efficiency improvements in doubly resonant cavities by simultaneously confining both fundamental and second-harmonic modes.[19-21] However, practical implementation confronts three fundamental challenges: i) out-of-plane mode leakage at SH wavelengths due to light cone limitations, ii) spatial mode mismatch between fundamental (FH) and second-harmonic (SH) modes, and iii) thermal instability under high-power operation. These issues have hindered the development of reliable on-chip light sources.

Recent advances in single-resonant silicon-carbide nanocavities with ultrahigh Q-factor have circumvented dual-resonance complexities, nabling direct SHG through precise control of coherence lengths and modal volumes.[22-25] Building on these progresses, we demonstrate a high-Q single-resonant photonic crystal nanobeam cavity (PCNBC) on a polymer-loaded lithium niobate on insulator (LNOI) platform. By optimizing lateral confinement and longitudinal refractive index modulation, our design achieves a visible-wavelength single-lobe SH emission at 768.77 nm with a normalized conversion efficiency of 163%/W, which is more than three orders of magnitude greater than previously reported LN-based photonic crystal cavities.[26,27] Direct CCD imaging of high-order cavity modes confirms strong spatial overlap between FH and SH modes. Furthermore, the device exhibits thermal stability under variations up to 20 °C, overcoming the thermal drift limitations of multi-resonant systems. This work provides a scalable architecture for efficient nonlinear light generation, with applications in quantum optics and chip-scale interconnects.

■ **Result and Discussion**

The cavity structure employed in this study, as described in Ref.[28], was fabricated by patterning a 400-nm-thick polymer layer on a y-cut LNOI substrate (NanoLN Inc.), comprising a 300-nm-thick LN layer and a 2-μm-thick buried oxide ($SiO_2$) layer. To achieve high-efficiency SHG on this platform, a TE-polarized PCNBC design and x-axis orientation of optical path were carefully selected. These conditions ensure that the dominant optical field of the TE mode is tightly confined within the LN layer, allowing the largest nonlinear component, $d_{33}$, to fully contribute to the nonlinear conversion. Building upon these principles, we designed and fabricated the devices with optimized lattice parameters. Key advancements over prior PCNBC designs include a reduced lattice constant of 432 nm for visible-band SHG and 140 polymer stripes in both taper and mirror regions to balance optical transmission and quality factor. Figure 1a shows scanning electron microscopy (SEM) images of the device, featuring grating couplers for input/output coupling.

Figure 1b illustrates the dual-band (telecom/visible) experimental setup. A tunable semiconductor laser (TSL, Santec/TSL550) with manual fiber polarization control (FPC) delivered TE-polarized pump light to the cavity. Transmitted telecom light was monitored using an InGaAs photodetector (IR PD, Thorlabs/S122C) and power meter (PM, Thorlabs/PM100). The insertion loss of grating couplers is estimated to be 10.5 dB/facet. The second-harmonic (SH) emission was collected via a 200-μm-core multimode fiber, which accommodates the mode profiles of the fundamental, second-order, and third-order modes. A 7° fiber tilt minimized reflection loss, while a spectrometer (Princeton Instruments/Acton Series SP-2500) and silicon photodetector (Thorlabs/S150C) analyzed SH spectra and power. Temperature stabilization was achieved using a thermoelectric cooler (TEC) with PID-controlled feedback from an embedded sensor.

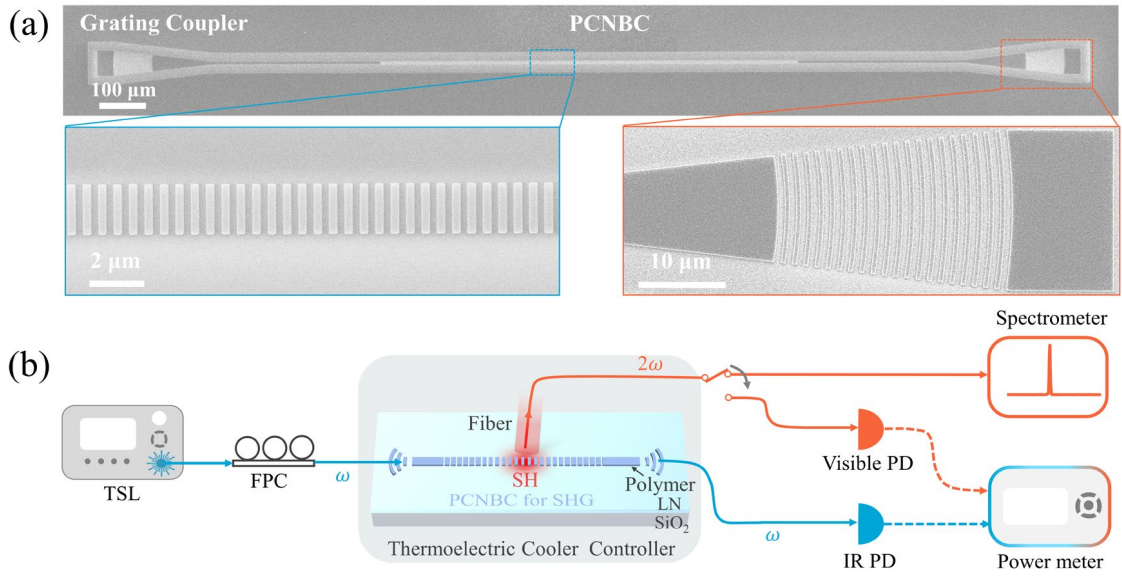

Figure 1. (a) SEM images of the fabricated PCNBC with grating couplers. (b) SHG characterization setup. TSL: tunable semiconductor laser; FPC: fiber polarization controller; PD: photodetector.

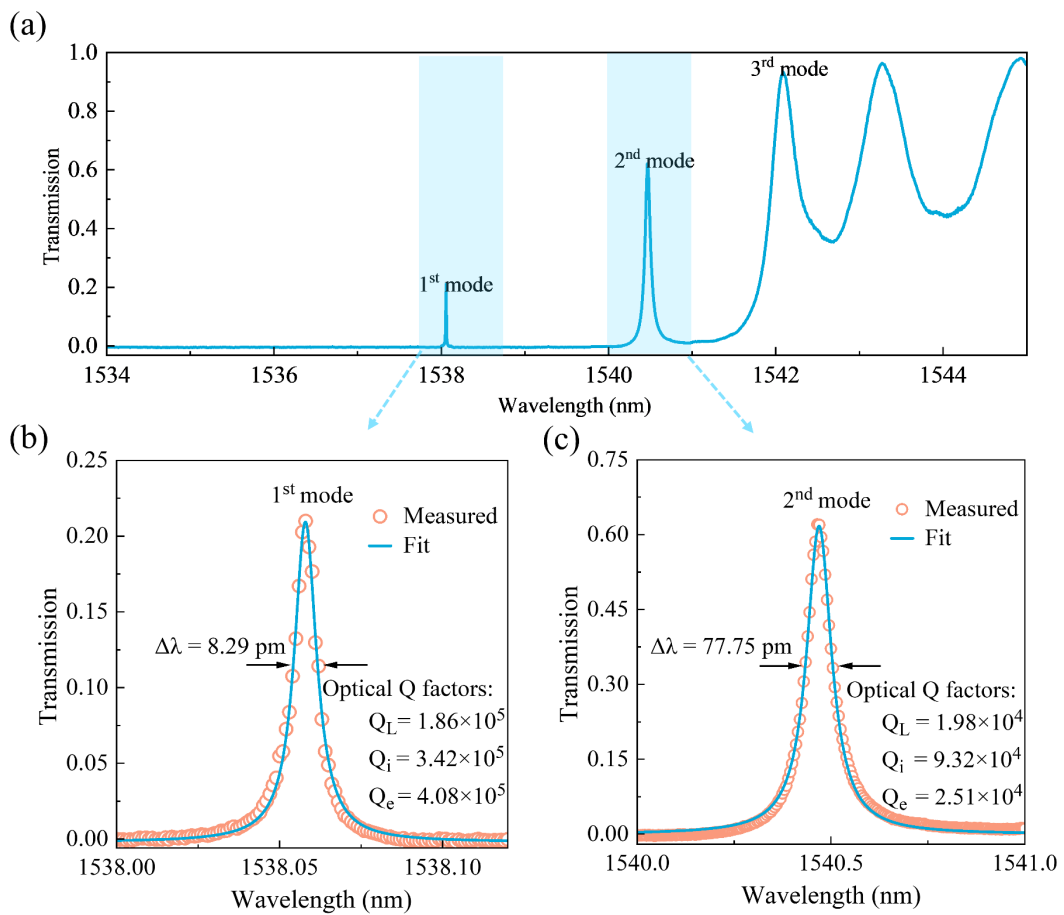

Figure 2. (a) Measured optical transmission spectrum of the PCNBC across a broad telecom band range. (b), (c) Detailed transmission spectra of the fundamental mode (b) and the second-order mode (c) show a loaded Q factor of $1.86×10^5$ around 1538.06 nm and $1.98×10^4$ around 1540.47 nm, obtained via Lorentzian fit. The extracted intrinsic Q factor ($Q_i$) and external coupling Q factor ($Q_e$) for the fundamental mode are $3.42×10^5$ and $4.08×10^5$, respectively, while for the second-order mode, they are $9.32×10^4$ and $2.51×10^4$.

As presented in Figure 2a, we first characterized the optical properties of the PCNBC in the telecom band by scanning the TSL wavelength from 1534.00 nm to 1545.00 nm. Resonances include fundamental mode at 1538.06 nm, second-order mode at 1540.47 nm, third-order mode at 1542.09 nm with spectra widening for higher modes due to increased losses. Figure 2b present the detailed transmission spectrum of the fundamental mode with a loaded Q factor ($Q_L$) of $1.86×10^5$ and a high transmission of 21%. Figure 2c present the transmission spectrum of the second-order mode with a higher transmission of 62% but a lower $Q_L$ of $1.98×10^4$. These performance metrics are better than the most PCNBCs reported to date.[29-31]

Under high-power excitation at the fundamental mode resonance, a distinct single-lobe emission pattern was observed at the nanocavity center. CCD imaging under dark-field conditions confirmed spatial confinement consistent with simulated mode profiles. Pumping higher-order modes generated multi-lobe patterns, marking the first observation of mode-dependent SH emission profiles in PCNBCs.[28,32] To confirm that the visible emissions correspond to SH generation, we fixed the laser wavelength at PCNBC resonances where visible emission were brightest. As shown Figure 3b, spectrometer measurements revealed SH peaks at 768.77 nm, 769.95 nm, and 770.84 nm, closely matching half the resonant wavelengths of the fundamental, second-order, and third-order modes. Minor deviations are attributed to the resolution limits of the spectrometer.

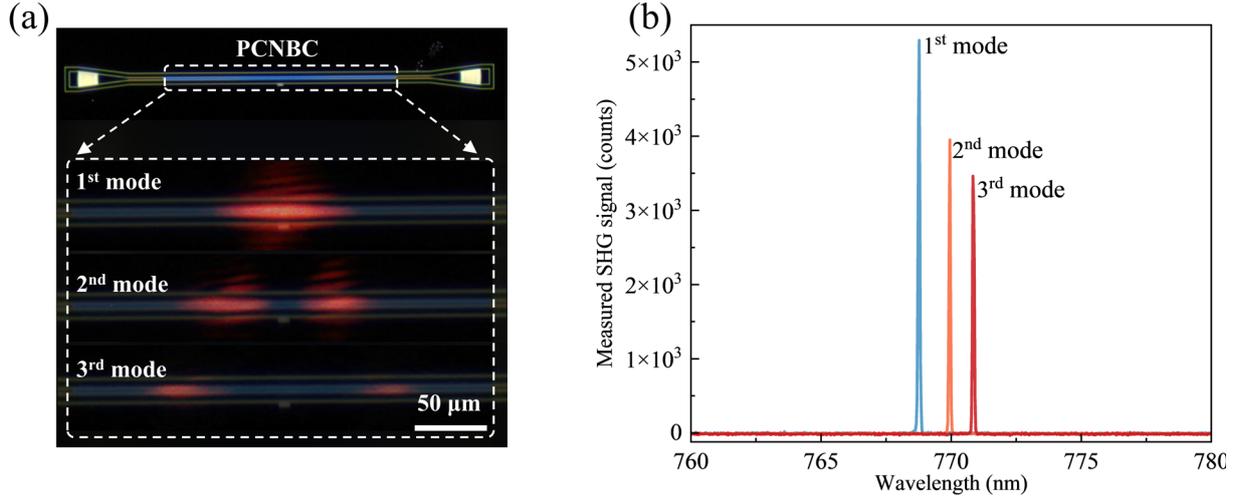

Figure 3. (a) SH emission patterns for fundamental, second-order, and third-order modes, respectively. (b) SH spectra recorded at fixed laser wavelengths for each resonant mode.

Using classical temporal coupled-mode theory,[33] we estimate the intracavity pump power ($P_{cav}$) contributing to SHG conversion. In the nondepleted (low pump power) regime, the intracavity field of the resonance mode is unaffected by the SHG process. Thus, in steady state, $P_{cav}$ can be expressed as

$$P_{cav} = \eta_{cav} P_{in} = \frac{2\tau_t^2}{\tau_i \tau_e} P_{in} \qquad (1)$$

where $\tau_t$, $\tau_i$ and $\tau_e$ represent the total, intrinsic, and external lifetimes of the resonance mode ($\tau_t^{-1} = \tau_i^{-1} + \tau_e^{-1}$). $P_{in}$ is the pump power after the grating coupler. Combining transmission spectra with the extracted Q factors in Figure 2, we calculate $\eta_{cav}$ for each resonance mode. The normalized SHG conversion efficiency η can be defined as $\eta = \frac{P_{SHG}}{P_{cav}^2}$.

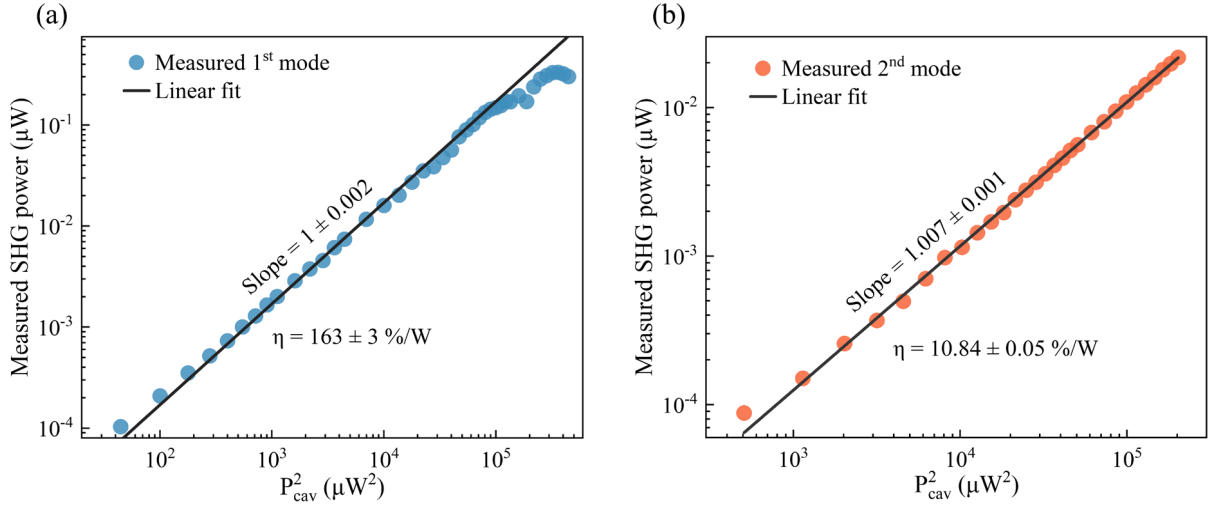

Figure 4. (a), (b) SHG power ($P_{SHG}$) versus the squared intracavity pump power ($P_{cav}^2$) for the fundamental mode (a) and the second-order mode (b). The sloid line represents a linear fit in the nondepleted region, with a normalized SHG conversion efficiency of 163%/W for the fundamental mode and 10.84%/W for the second-order mode.

Figure 4a shows the SHG power $P_{SHG}$ versus $P_{cav}^2$ for the fundamental mode. In the non-pump depleted regime ($P_{cav}$<400 μW), the linear fit (slope=1) confirms the expected quadratic dependence, yielding a normalized conversion efficiency of 163%/W with an uncertainty of 3%/W. At higher powers ($P_{cav}$>400 μW), deviations from quadratic scaling indicate pump depletion, with the absolute efficiency reaching 0.055% at $P_{cav}$ of 601.72 μW. For the second-order mode, the normalized efficiency drops to 10.84%/W due to its lower Q factor.[24]

Table 1 benchmarks SHG efficiencies across photonic crystal cavities platforms. While SiC cavities achieve higher efficiencies, their SHG emission patterns deviate spatially from the FH modes, which would introduce challenges in subsequent processes such as light source collimation, fiber transmission, and inter-chip coupling.[34-36] In contrast, our LN-based PCNBC achieves a normalized efficiency of 163%/W, which is more than three orders of magnitude higher than the other LN PhC cavities. In addition to its high SHG conversion efficiency, an important feature of our device is that the generated SH emission perfectly matches the fundamental harmonic modes, making it particularly suitable as a light source for photonic integrated circuits (PICs).

Table 1 Comparison of SHG conversion efficiencies for photonic crystal cavities

| Cavity type/Platform | $Q_L$ (×10$^4$) | SH collection method | η (%/W) | SH mode consistent with FH mode |
|---|---|---|---|---|
| 2D PhC/SiC[23] | 60 | Microscope objective | 1900 | No |
| 2D PhC/SiC[24] | 71 | Microscope objective | 4000 | No |
| 2D PhC/GaP[37] | 0.56 | Microscope objective | 430 | No |
| 2D PhC/GaAs[38] | 0.4 | Microscope objective | 1.2 | No |
| 2D PhC/LN[27] | 0.05 | Microscope objective | $1.2×10^{-2}$ | No |
| 2D PhC/LN[26] | 33.4 | Tapered optical fiber | $7.8×10^{-2}$ | Yes |
| PCNBC/LN[39] | 5.4 | Tapered optical fiber | $0.4×10^{-3}$ | Yes |
| PCNBC/LN[This work] | 18.6 | Multimode fiber | 163 | Yes |

As illustrated in Figure 5a, thermal tuning tests reveal a redshift in both pump and SH wavelengths with increasing temperature, consistent with the thermo-optic effect. The SH wavelength is recorded via the spectrometer when the laser wavelength is fixed at the resonance wavelength of the fundamental mode, with a low $P_{cav}$ of 34 µW. The pump wavelength shifts with a linear fitted temperature tuning rate of 26.8±0.23 pm/°C is roughly twice that of the SH of 13.5±0.37 pm/°C. Crucially, as shown in Figure 5b, SHG conversion efficiency remains stable across a 20°C range, demonstrating robustness against thermal drift which is a key advantage over multi-resonant or phase match SHG devices.[40,41]

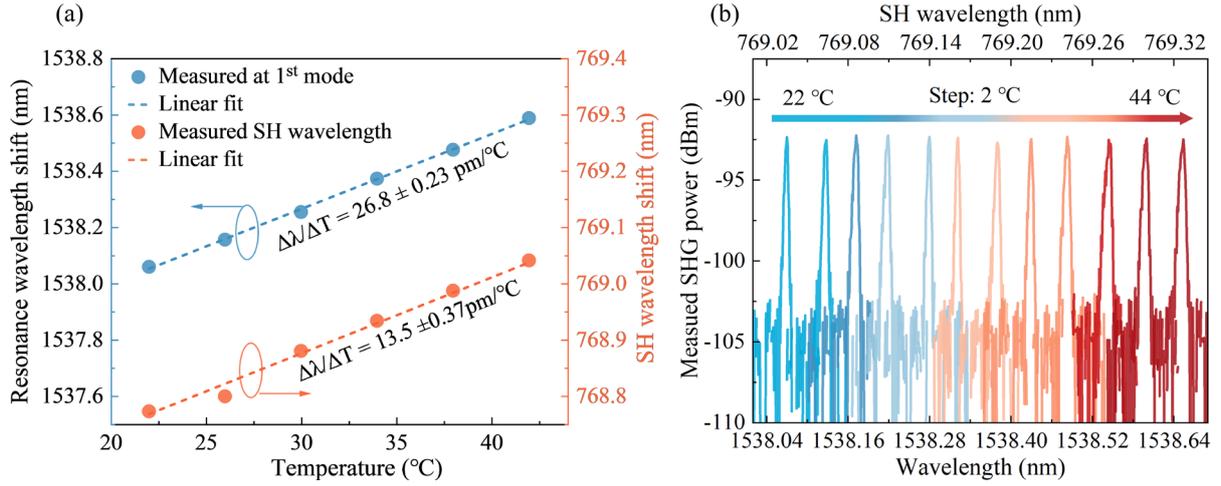

Figure 5. (a) Resonance wavelength shifts for pump (fundamental mode) and SH light versus temperature. (b) SH spectra show consistent peak power at different temperatures.

## ■ CONCLUSIONS

In conclusion, we have proposed and experimentally demonstrated a high-Q single-resonant nanocavity for efficient SHG on polymer-loaded y-cut LNOI platform. The device emits a bright single-lobe radiation pattern at 768.77 nm, achieving a normalized SHG conversion efficiency of 163%/W, which is more than three orders of magnitude higher than that of previous similar microcavities. In addition, the devices exhibit exceptional temperature stability, with no significant significant efficiency degradation over a 20 °C range. Importantly, this approach is not limited to the LN material, but may also be universally applied to other photonics platforms, such as barium titanate (BTO), aluminum nitrogen (AlN), particularly for materials that are difficult to be etched. Our work establishes a scalable architecture for ultracompact nonlinear light sources, with applications in quantum optics and chip-scale interconnects.

## AUTHOR INFORMATION

**Corresponding Authors**

**Danyang Yao** – State Key Laboratory of Wide-Bandgap Semiconductor Devices and Integrated Technology, School of Microelectronics, Faculty of integrated circuits, Xidian University, Xi'an 710071, China; Email: dyyao@xidian.edu.cn;

**Xuetao Gan** – Key Laboratory of Light Field Manipulation and Information Acquisition, Ministry of Industry and Information Technology, and Shaanxi Key Laboratory of Optical Information Technology, School of Physical Science and Technology, Northwestern Polytechnical University, Xi'an, 710129, China; Email: xuetaogan@nwpu.edu.cn;

**Jinchuan Zhang** – Laboratory of Solid-State Optoelectronics Information Technology, Institute of Semiconductors, Chinese Academy of Sciences, Beijing 100083, China; Email: zhangjinchuan@semi.ac.cn;

**Authors**

**Zhi Jiang** – State Key Laboratory of Wide-Bandgap Semiconductor Devices and Integrated Technology, School of Microelectronics, Faculty of integrated circuits, Xidian University, Xi'an 710071, China

**Yu Gao** – State Key Laboratory of Wide-Bandgap Semiconductor Devices and Integrated Technology, School of Microelectronics, Faculty of integrated circuits, Xidian University, Xi'an 710071, China

**Xu Ran** – State Key Laboratory of Wide-Bandgap Semiconductor Devices and Integrated Technology, School of Microelectronics, Faculty of integrated circuits, Xidian University, Xi'an 710071, China

**Duomao Li** – State Key Laboratory of Wide-Bandgap Semiconductor Devices and Integrated Technology, School of Microelectronics, Faculty of integrated circuits, Xidian University, Xi'an 710071, China

**Erqi Zhang** – State Key Laboratory of Wide-Bandgap Semiconductor Devices and Integrated Technology, School of Microelectronics, Faculty of integrated circuits, Xidian University, Xi'an 710071, China

**Jianguo Wang** – Key Laboratory of Light Field Manipulation and Information Acquisition, Ministry of Industry and Information Technology, and Shaanxi Key Laboratory of Optical Information Technology, School of Physical Science and Technology, Northwestern Polytechnical University, Xi'an, 710129, China

**Fengqi Liu** - Laboratory of Solid-State Optoelectronics Information Technology, Institute of Semiconductors, Chinese Academy of Sciences, Beijing 100083, China

**Yue Hao** – State Key Laboratory of Wide-Bandgap Semiconductor Devices and Integrated Technology, School of Microelectronics, Faculty of integrated circuits, Xidian University, Xi'an 710071, China

**Notes**

The authors declare no competing financial interest.

**ACKNOWLEDGMENT**

This work was supported by the Chinese Institute of Electronics-Smartchip Special Research Program, the National Natural Science Foundation of China (Grant No. 62293522 and 62235016), and the Fundamental Research Funds for the Central Universities.